\documentstyle[twocolumn,aps,prd,epsf,epsfig]{revtex}
\def\cosp{\sin (\nu +\beta )} 
\def\sinp{\cos (\nu +\beta )} 

\begin{document}

\twocolumn[\hsize\textwidth\columnwidth\hsize
\csname@twocolumnfalse%
\endcsname
\draft
\title{Adiabatic pulse propagation in coherent atomic media with the 
tripod level
 configuration}
\author{I.E.~Mazets}
\address{{\setlength{\baselineskip}{18pt}
Ioffe Physico-Technical
Institute, 194021 St.Petersburg, Russia, \\ 
Instutut f\"{u}r Experimentalphysik, TU Graz, 8010 Graz, Austria}
} 
\maketitle

\begin{abstract}
We investigate the problem of propagation of
 three-component resonant 
light pulses with adiabatically varying amplitudes
through a medium consisting of atoms with the tripod level configuration. 
By means of both analytic and numerical methods we found the two 
modes of shape-preserving pulse propagation. 
The
 pulse propagation velocity is found to be either equal to the speed of
light or significantly slowed down, depending on a particular propagation 
mode. 
\\  
\end{abstract}
\vskip1pc]

\section{Introduction}

Coherent population trapping (CPT) is a well-known phenomenon of preparation 
of atoms in a coherent superposition of ground or metastable state sublevels
(so-called dark state), which is immune to excitation by a two-component
laser radiation under the two-photon resonance condition 
\cite{arimrev}. Since the laser radiation is not scattered
by atoms in the dark state, the radiation absorption is dramatically
reduced. This effect is called electromagnetically induced transparency
(EIT) and is actively studied since early 90's 
\cite{harrev}. One of the most striking features of EIT is possibility of
shape-preserving propagation of light pulses with slowly (adiabatically)
varying amplitudes at the group velocity significantly reduced with respect
to the speed of light in vacuum, $c$ \cite{adiabaton}. Shape-preserving 
electromagnetic pulses propagating in a coherent atomic medium at 
the reduced group velocity were called in Ref.~\cite{adiabaton} 
``adiabatons''. Experimental
observations of light a pulse group velocity less by many orders of
magnitude than $c$ has been repeatedly reported \cite{glc}. Slowing down the
laser light followed by spatial compression of the pulses provides a unique
possibility for design of nonlinear-optical devices operating on a
few-photon level \cite{harhau}. Extreme sensitivity of CPT and EIT to
deviations from the two-photon resonance allowed to observe experimentally
large Kerr nonlinearity \cite{kerr} and absorptive optical switching 
\cite{switch} in cold rubidium vapor. 
Such nonlinear optical phenomena, along
with the possibility of reversible conversion of a photonic excitation to a
collective spin excitation \cite{lrmp} and trapping light in a medium with
the photonic band gap induced by a periodic modulation of the EIT resonance 
\cite{bgap}, are of great importance for quantum information storage and
processing.

A novel direction in CPT and EIT studies is related to the systems admitting
more than one dark state for the given real Rabi frequencies, 
$|\Omega _{j}|$,
and phases, $\chi _{j}$, associated with the resonantly driven transitions.
The simplest scheme of such a kind is the tripod scheme displayed in 
Fig.~1. 
Stimulated Raman adiabatic passage (STIRAP) in such an optically
thin atomic medium with the tripod level scheme was investigated
theoretically \cite{bergth} and demonstrated experimentally \cite{bergex} by
Bergmann and co-workers. A proposal to use the tripod scheme as a physical
implementation of a qubit has been made recently \cite{renzoni}. A 5-level
scheme being the extension of the tripod scheme was considered in 
Ref.~\cite{stenholm}.

\begin{figure}[h]
\begin{center} 
\centerline{\epsfig{file=bt1.ps,width=5cm}}
\end{center} 
\vspace{-0.4cm} 
\begin{caption}
{Tripod scheme of levels driven by resonant electromagnetic fields.}
\end{caption}
\label{fig1}
\end{figure}

The very specifics of the tripod scheme is that during adiabatically slow
change of the external field parameters transitions between the two dark
state occurs. These transitions are described by a non-Abelian phase matrix 
\cite{wilcz}, which is a generalization of a geometric (Berry) phase 
\cite{berry} 
to the case of degenerate eigenstates of an adiabatic Hamiltonian.
One may expect that these transitions give rise to rich and complicated
dynamics of laser pulse propagation in an optically dense medium with the
tripod level configuration. However, only few theoretical works on EIT in
such media are available. Paspalakis and Knight \cite{pk} considered
parametric frequency generation for the case of time-independent fields at
the medium entrance and calculated the group velocity of a weak probe field.
Petrosyan and Malakyan \cite{pm} investigated theoretically EIT in a tripod
medium as a tool for optical cross-phase modulation and high-precision
magnetometry in the weak probe field limit. The value of the group velocity
obtained in \cite{pk,pm} is strongly reduced with respect to $c$ in the same
way as in the standard case of EIT in a three-level medium \cite{adiabaton}.
In the theoretical interpretation of the experiment on four-wave mixing in a
solid-state system with the tripod level configuration \cite{hamb} and other
numerical calculations by Ham \cite{hama} related to that system, small
optical density of the medium was assumed.

The aim of the present paper is to study pulse propagation in a 
medium with the tripod level scheme (hereafter briefly called ``tripod 
medium'') for a general case, in which none of the three resonant 
electromagnetic fields is assumed to be weak compared to others. 
The paper is organized as follows. 
In Sec.~II we  present the set of equations
treating pulse propagation in the adiabatic regime in a tripod medium 
beyond
 the weak probe approximation
. In Sec.~III the analytic solutions 
describing slow and fast pulse propagation are obtained. Sec.~IV 
contains the results of our numerical calculations and their 
interpretation. Sec.~V deals with some particular regimes of 
propagations. Sec.~VI is devoted to conclusive remarks. 

\section{Basic equations} 

If the three electromagnetic fields are tuned exactly in resonance with 
the corresponding transitions $\left| j \right \rangle \leftrightarrow 
\left| 0\right \rangle $, $j=1,2,3$, the Hamiltonian in the 
interaction representation reads as 
\begin{equation} 
\hat{H}=-\hbar \sum _{j=1}^3 \Omega _j \left| 0 
\right \rangle \left \langle j\right| +\mathrm{H.c.}, 
\label{hmt} 
\end{equation} 
$\Omega _j\equiv |\Omega _j|e^{i\chi _j}
=d_{0j}E_j/\hbar $, where $d_{0j}$ is the dipole moment 
matrix element of the given transition. The electric field in the 
$j$th laser wave is $E_j\exp [ik_j(z-t/c)]+\mathrm{c.c.}$, $k_j$ being 
the radiation wave number. The complex 
amplitude $E_j$ is a slowly varying function of $z$ and $t$. 
Expanding the atomic wave function as $\left| \psi \right \rangle =
a_0\left| 0\right \rangle + \sum _{j=1}^3a_j\left| j\right \rangle $, 
we obtain the Schr\"{o}dinger equation for the probability 
amplitudes: 
\begin{eqnarray} 
i\dot{a}_j&=&-\Omega _j^*a_0, \quad j=1,2,3,\nonumber \\
i\dot{a}_0&=&-\sum _{j=1}^3\Omega _ja_j. \label{schr} 
\end{eqnarray} 
The set of shortened Maxwell equations for slowly varying field 
amplitudes can be written as 
\begin{equation} 
\left( \frac \partial {\partial z}+\frac 1c \frac \partial {\partial t}
\right) \Omega _j =iG_j a_0a_j^*, \quad j=1,2,3, 
\label{mxw} 
\end{equation} 
where $G_j=2\pi k_j n d_{0j}^2/\hbar $ and $n$ is the atomic number 
density. Taking into consideration propagation effects described by 
Eq.~({\ref{mxw}) is the essence of the theory developed in the present 
Section, in contrast to the theory of Ref.~\cite{bergth}, which 
applies to the case of a refractively thin medium. 

Hereafter we assume that all the matter-field coupling 
constants are equal: 
\begin{equation} 
G_1=G_2=G_3\equiv G. \label{allG}
\end{equation} 
Violation of this assumption leads to adiabaticity breakdown 
during the pulse propagation and subsequent pulse front 
steepening \cite{neos}. Thermal motion of atoms leads, 
besides reduction of the 
effective number density of atoms in resonance with the laser radiation, 
to a similar effect of pulse front steepening \cite{ds}. However, the 
pulse shape distortion effects manifest themselves at propagation 
distances much larger than the typical propagation distance associated 
with an adiabaton-like pulse formation \cite{neos,ds}. 
Therefore we can neglect 
both the differences of oscillator strengths associated with the 
three laser-driven transitions and thermal motion of atoms. 
We also neglect the radiative decay of the excited state 
$\left| 0\right \rangle $ since it plays no role in the adiabatic 
regime, because of negligible population of the excited state 
\cite{adiabaton}. 

We parameterize the Rabi frequencies by introducing the generalized 
Rabi frequency $\Omega =\left( \sum _{j=1}^3 |\Omega _j|^2 \right) 
^{1/2}$ and two angular variables $\varphi $ and $\vartheta $: 
\begin{eqnarray} 
\Omega _1&=&\sin \vartheta \cos \varphi \, e^{i\chi _1}\Omega , 
\nonumber \\
\Omega _2&=&\cos \vartheta \cos \varphi \, e^{i\chi _2}\Omega , 
\label{para} \\
\Omega _3&=&\sin \varphi \, e^{i\chi _3}\Omega . \nonumber 
\end{eqnarray} 
There are two mutually orthogonal 
non-absorbing (dark) states associated with the 
Hamiltonian of Eq.~(\ref{hmt}): 
\begin{eqnarray} 
\left| \Phi ^{(1)} \right \rangle &=& \cos \vartheta \, e^{-i\chi _1} 
\left| 1\right \rangle -\sin \vartheta \,e^{-i\chi _2}\left| 2 
\right \rangle ,\nonumber \\
\left| \Phi ^{(2)} \right \rangle &=& \sin \vartheta \sin \varphi \, 
e^{-i\chi _1} \left| 1\right \rangle + \label{darkst} \\
&& \cos \vartheta \sin \varphi \, e^{-i\chi _2} \left| 2\right \rangle -
\cos \varphi \, e^{-i\chi _3}\left| 3\right \rangle . \nonumber 
\end{eqnarray} 
An atom initially prepared in either of these two states remains 
unexcited since 
\begin{equation} 
\hat{H}\left| \Phi ^{(s)} \right \rangle =0, \quad s=1,2. 
\label{prodark} 
\end{equation}
An atom also remains unexcited if the parameters of the laser 
radiation vary in time slowly enough to satisfy the adiabaticity 
conditions 
\begin{equation} 
\dot{\vartheta }\ll \Omega , ~\qquad \dot{\varphi }\ll \Omega , 
\label{ac1} 
\end{equation} 
and 
\begin{equation} 
\dot{\chi }_j\ll \Omega ,~\quad j=1,2,3.     \label{ac2} 
\end{equation} 
However, in the latter case there are 
transitions \cite{wilcz} between the dark states 
defined by Eqs.~(\ref{darkst}), where the instantaneous values of 
the varying angles $\varphi $ and $\vartheta $ and phases $\chi _j$ 
enter. If at $t\rightarrow -\infty $ an atom was in the $s$th dark 
state, its wave function at subsequent instants of time is 
\begin{equation} 
\left| \Psi ^{(s)}\right \rangle =\sum _{s^\prime =1}^2 
B_{ss^\prime }(t)\left| \Phi ^{(s^\prime )} \right \rangle , 
\label{psis}
\end{equation}
where the matrix $\hat{B}$ obeys the equation 
\begin{equation} 
\dot{B}_{ss^\prime }(t)+\sum _{s^{\prime \prime }}
B_{ss^{\prime \prime }}(t)A_{s^\prime s^{\prime \prime }}(t)=0 
\label{evB}
\end{equation} 
with the initial condition 
\begin{equation} 
B_{ss^\prime }(-\infty )=\delta _{ss^\prime },    \label{inB} 
\end{equation}
and 
\begin{equation} 
A_{s^\prime s^{\prime \prime }}(t) =\left \langle \Phi ^{(s^\prime )}
\right| \frac \partial {\partial t} 
\left| {\Phi }^{(s^{\prime \prime })} \right \rangle . 
\label{genA}
\end{equation} 
Explicitly, 
\begin{eqnarray}
A_{11}&=&
-i(\dot{\chi }_1\cos ^2\vartheta +\dot{\chi }_2\sin ^2\vartheta ), 
\nonumber \\ 
A_{12}&=&\dot{\vartheta }\sin \varphi -  
i(\dot{\chi }_1-\dot{\chi }_2)\sin \vartheta \cos \vartheta 
\sin \varphi  ,          \nonumber \\
A_{21}&=&-A_{12}^*   ,   \nonumber \\
A_{22}&=&-i[(\dot{\chi }_1\sin ^2\vartheta +
\dot{\chi }_2\cos ^2\vartheta )\sin ^2\varphi +  \nonumber \\
&&\dot{\chi }_3 \cos ^2\varphi ].      \label{pA1} 
\end{eqnarray}

It is easy to show that if the phases of the laser fields are kept 
constant at the medium entrance, then 
\begin{equation} 
\dot{\chi }_j=0, \quad j=1,2,3,   \label{npv} 
\end{equation} 
in the whole tripod medium. The opposite is not true. If the 
absolute values of the field amplitudes are constant at the 
medium entrance, but the phases are modulated, then the absolute 
values of the fields amplitudes and, hence, $\vartheta $ and 
$\varphi $ become time-dependent inside the medium. In the present 
paper we consider only the case when Eq.~(\ref{npv}) holds. 
In this case Eqs.~(\ref{pA1}) are reduced to 
\begin{equation} 
A_{11}=A_{22}=0,   \qquad
A_{12}=-A_{21}=\dot{\nu}   ,    \label{pA2}
\end{equation} 
where 
\begin{equation} 
\dot{\nu }=\dot{\vartheta }\sin \varphi  
\label{basenu} 
\end{equation} 
and $\nu (z,-\infty )=0$. Then Eqs.~(\ref{evB}, \ref{inB}) yield the 
following result \cite{bergth}:  
\begin{equation} 
B_{11}=B_{22}=\cos \nu ,   \qquad 
B_{12}=-B_{21}=\sin \nu   .    \label{B2nu}
\end{equation} 

We assume that the tripod medium occupies the half-space $z>0$. 
Initially, at $t\rightarrow -\infty $, all the atoms in the medium 
are in the coherent superposition of the dark states 
\begin{equation} 
\left| \psi (-\infty  )\right \rangle =\cos \beta 
\left| \Phi ^{(1)}\right \rangle +\sin \beta 
\left| \Phi ^{(2)}\right \rangle . 
\label{inpsi}
\end{equation} 
The boundary conditions for the fields at the medium entrance 
$\Omega (0,t)=\Omega _0(t)$, $\vartheta (0,t)=
\vartheta _0(t)$, and $\varphi (0,t)=\varphi _0(t)$ are consistent 
with Eq.~(\ref{ac1}). Thus the adiabatic regime of the 
laser radiation propagation inside the medium is ensured. 
It is convenient to introduce new variables $\zeta =z$ and 
$\tau =t-z/c$, as in Ref.~\cite{adiabaton}. Respectively, the 
derivatives over the new variables are $\partial /(\partial \tau )= 
\partial /(\partial t )$ and $\partial /(\partial \zeta )=
\partial /(\partial z )+c^{-1}\partial /(\partial t )$. 

Now we can solve self-consistently the set of Schr\"{o}dinger --- Maxwell 
equations (\ref{schr},~\ref{mxw}). First of all, we note that in the 
adiabatic regime $a_0$ is very small, and the probability 
amplitudes of the low-energy states ($j=1,2,3$) 
are, according to Eqs.~(\ref{psis}, 
\ref{B2nu}, \ref{inpsi}),
\begin{equation}
a_j=\sinp \left \langle j\right| \Phi ^{(1)}\left. \right 
\rangle +  
\cosp \left \langle j\right| \Phi ^{(2)}\left. \right \rangle . 
\label{ajbn}
\end{equation} 
Then we find easily, that, similarly to the case of adiabatic pulse 
propagation in a $\Lambda $-medium \cite{adiabaton}, 
\begin{equation} 
\frac \partial {\partial \zeta }\Omega =0,   \label{omtau} 
\end{equation}
i.e., $\Omega =\Omega _0(\tau )$. Then we use the trick first applied 
in Ref.~\cite{adiabaton}: We express the small probability amplitude 
of the excited state as $a_0= -(i/\Omega _j^*)\partial a_j/
(\partial \tau )$ and substitute this expression 
into the shortened Maxwell equations (\ref{mxw}). We get 
$\partial |\Omega _j/\Omega |^2/(\partial \zeta )=
(G/\Omega ^2)\partial |a_j|^2/(\partial \tau )$, $j=1,2,3$, or, 
explicitly, 
\begin{eqnarray}
\frac \partial {\partial \zeta }(\sin \vartheta \cos \varphi )^2&=&
\frac \partial {\partial w} [\sinp \cos \vartheta + \nonumber \\ && 
\cosp \sin \vartheta \sin \varphi ]^2, \nonumber \\
\frac \partial {\partial \zeta }(\cos \vartheta \cos \varphi )^2&=&
\frac \partial {\partial w} [-\sinp \sin \vartheta + \nonumber \\ &&
\cosp \cos \vartheta \sin \varphi ]^2, \nonumber \\ 
\frac \partial {\partial \zeta }\sin ^2 \varphi &=&
\frac \partial {\partial w}[\cosp \cos \varphi ]^2  . \label{trzve}
\end{eqnarray} 
Here we introduced, instead of $\tau $, a new variable (nonlinear time) 
\begin{equation} 
w=\frac 1G \int _{-\infty }^\tau \Omega _0^2(\tau ^\prime ) \, 
d\tau ^\prime ,       \label{defw} 
\end{equation}
which has the dimension of length. Then Eq.~({\ref{basenu}) 
takes the form 
\begin{equation} 
\frac {\partial \nu }{\partial w} =\frac {\partial 
\vartheta }{\partial w}\sin \varphi .         \label{usenu} 
\end{equation} 
All the initial conditions set at $t\rightarrow -\infty $ apply now 
to $w=0$. 

Only two of Eqs. (\ref{trzve}) are independent. After some tedious 
calculations they are reduced to 
\begin{eqnarray} 
\frac {\cosp }{\cos \varphi } \left( \frac \partial {\partial \zeta }+
\frac \partial {\partial w }\right) \varphi -&& \nonumber \\ 
\sinp \left( \frac \partial {\partial \zeta }+
\frac \partial {\partial w }\right) \vartheta &=&0,  \nonumber \\
\frac {\sinp }{\cos \varphi }\frac {\partial \varphi }{\partial \zeta }
+\cosp \frac {\partial \vartheta }{\partial \zeta }&=&0. 
\label{workn}
\end{eqnarray}

It is convenient now to change the variables to $u_1=\zeta -w$ 
and $u_2=w$. The set of Eqs.~(\ref{usenu},~\ref{workn}) takes the 
form 
\begin{eqnarray} 
\frac {\sinp }{\cos \varphi }\frac {\partial \varphi }{\partial u_1 }
&+&\cosp \frac {\partial \vartheta }{\partial u_1}=0, \label{wa1} \\
\frac {\cosp }{\cos \varphi }\frac {\partial \varphi }{\partial u_2 }
&-&\sinp \frac {\partial \vartheta }{\partial u_2}=0, \label{wa2} \\
\frac {\partial \nu }{\partial u_1}-\frac {\partial \nu }{\partial u_2} 
&=& \left( \frac {\partial \vartheta }{\partial u_1}-
\frac {\partial \vartheta }{\partial u_2} 
\right) \sin \varphi .            \label{wa3}   
\end{eqnarray}  

\section{Slow and fast pulses: The analytic solution}

The set of Eqs.~(\ref{wa1} -- \ref{wa3}) is especially convenient for 
searching analytic solutions in a case when the unknown functions 
$\varphi $ and $\vartheta $ depend on only one of the variables 
$u_1,~u_2$. We find two classes of solutions. 

The first one is the class of slow pulses. In this case the unknown 
functions depend only on $u_1=\zeta -w$. The group velocity $v_g$ of 
pulses of such has the same form as that of adiabatons in a 
$\Lambda $-medium \cite{adiabaton}: $v_g=\left( c^{-1} +G/\Omega ^2 
\right ) ^{-1}$ and can be much less than $c$. All the derivatives over 
$u_2$ vanish, thus making Eq.~(\ref{wa2}) an identity. The two 
remaining equations (\ref{wa1}) and (\ref{wa3}) become ordinary 
differential equations, yielding the general solution in the 
parametric form: 
\begin{equation} 
|\cos \varphi | =\frac {C_1}{|\sinp |},  \quad 
|\sin (\vartheta -C_2)|=\frac {|\cosp |}{\sqrt{1-C_1^2}}. 
\label{slow1}
\end{equation}
Here $C_1,~C_2$ are arbitrary constants, and $p$ is any function 
of $\zeta -w$ compatible with the adiabaticity conditions (\ref{ac1}). 

Similarly, we find a general solution for the class of fast pulses, 
propagating at the speed of light: 
\begin{equation} 
|\cos \varphi | =\frac {C_3}{|\cosp |},  \quad 
|\sin (\vartheta -C_4)|=\frac {|\sinp |}{\sqrt{1-C_3^2}}. 
\label{fast1}
\end{equation}
Here $C_3,~C_4$ are arbitrary constants, and $p=p(w)$ must be 
compatible with Eq.~(\ref{ac1}). 

Although the set of Eqs.~(\ref{wa1} -- \ref{wa3}) looks rather 
simple and symmetric, our attempts to find its general solution in 
the case of dependence of $\varphi $ and $\vartheta $ on both $u_1$ 
and $u_2$ have been unsuccessful. However, we can prove that 
a time-dependent solution in the parametric form 
\begin{equation} 
\varphi =\varphi (p), \qquad \vartheta =\vartheta (p) 
\label{bdas}
\end{equation} 
does not exist if 
\begin{equation} 
\frac {\partial p}{\partial u_1}\neq 0,  \qquad 
\frac {\partial p}{\partial u_2}\neq 0.  \label{pu12}
\end{equation}
Indeed, Eqs.~(\ref{wa1}, \ref{wa2}) can be considered as linear 
homogeneous algebraic equations for $\cosp $ and $\sinp $. They 
have a solution if 
\begin{equation} 
\frac {\partial \vartheta }{\partial u_1}
\frac {\partial \vartheta }{\partial u_2}+\frac 1{\cos ^2\varphi }
\frac {\partial \varphi }{\partial u_1}
\frac {\partial \varphi }{\partial u_2}=0.          \label{sc0}
\end{equation}
But if we make an assumption given by Eq.~(\ref{bdas}) then 
Eq.~(\ref{sc0}) results in 
\begin{equation}
\frac {\partial p}{\partial u_1} 
\frac {\partial p}{\partial u_2}\left[ \left( 
\frac {d\vartheta }{dp}\right) ^2+\frac 1{\cos ^2 \varphi } \left( 
\frac {d\varphi }{dp}\right) ^2\right] =0.          \label{sc1}
\end{equation} 
If Eq. (\ref{pu12}) holds, it follows from Eq.~(\ref{sc1}) that 
$\varphi =\mathrm{const}$ and $\vartheta =\mathrm{const}$, 
i.e., there is no variation of the electromagnetic fields in 
space and time. 

The fact that we have not found other pulse group velocities than $c$ and 
$\left( c^{-1} +G/\Omega ^2 \right ) ^{-1}$ is in full agreement with the 
results of perturbative approach \cite{pk,pm}.  

\section{Numerical solutions}

It is natural to expect that any pulse of finite duration 
evolves in the medium into pair of fast and slow pulses, which 
become more and more separated in space due to the difference of 
their group velocities. Indeed, our numerical simulations 
confirm such an expectation. An example is shown in Fig.~2. 
The quantity $w_0$ used for normalization of the horizontal axes 
of the plots in Fig.~2 and subsequent determines the order of magnitude 
of $\dot{\varphi }$ and $\dot{\vartheta }$, which are $\sim \Omega ^2/
(Gw_0)$. The adiabaticity condition (\ref{ac1}) results in the following 
restriction: $w_0 \gg \Omega /G$. 
One can see that the incident pulse evolves into a 
well separated pair of fast (F) and slow (S) pulses, 
and the mixing angle $\nu $ describing transitions between the two 
dark states emerges (the incident pulse is chosen in such a form that 
$\nu \equiv 0$ at the medium entrance). The F and S pulses at 
large propagation distances can be excellently fitted with formulae 
(\ref{fast1}) and (\ref{slow1}), respectively. 

We also investigated numerically collisions between fast and slow 
pulses. The results are presented in Fig.~3. The pulse 
sequence is organized in such a way that the pulse of a shape 
satisfying Eq.~(\ref{slow1}) enters the medium first. After some 
time delay the next pulse obeying Eq.~(\ref{fast1}) enters the 
medium. The first pulses propagates at the slow group velocity 
whereas the second one propagates at the speed of light. The distance 
between them decreases, and at certain $\zeta $ the two pulses 
overlap (this is marked by O in Fig.~3b). 
Their nonlinear interaction leads to strong distortion of 
their shapes, which becomes apparent at larger propagation distances. 
Thus adiabatic pulses in a tripod medium cannot be called solitons 
in the exact sense of soliton definition by Zabusky and Kruskal 
\cite{soliton1}. Note that it is impossible to arrange a collision of 
two adiabatons in a $\Lambda $-medium. 

The parameter $\beta $ is equal to 1.12 for Fig.~2 and 1.87 
for Fig.~3. 

\begin{figure} 
\begin{center} 
\centerline{\epsfig{file=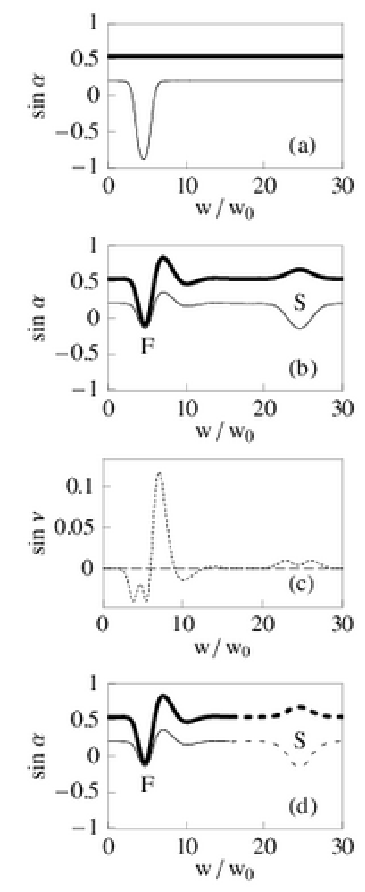,width=7cm}}
\end{center} 
\vspace{-0.6cm} 
\begin{caption} 
{Splitting of the incident pulse into the fast (F) and slow (S) pulses. 
(a) Sine functions of the angles $\alpha =\vartheta $ (thick line) 
and $\varphi $ (thin line) versus scaled nonlinear time, $w/w_0$, 
at the medium entrance, $\zeta =0$ (the boundary conditions); 
the same line styles are reserved for $\vartheta $ and $\varphi $ 
in all the subsequent figures. (b) The same as in (a) for $\zeta =20w_0$ 
(numerical simulation results). 
(c) Sine of mixing angle $\nu $ 
versus $w/w_0$ for $\zeta =0$ (dashed line) and $20w_0$ (dotted line). 
(d) Fitting of the plot (b) with Eq.~(\ref{fast1}) (solid lines;  
$C_1=0.88$, $C_2=1.75$) and Eq.~(\ref{slow1}) (dashed lines; $C_3=0.43$, 
$C_4=-0.90$. Axes are in dimensionless units in all plots.}
\end{caption} 
\label{fig2} 
\end{figure} 

\begin{figure} 
\begin{center} 
\centerline{\epsfig{file=bt3.ps,width=7cm}}
\end{center} 
\vspace{-0.6cm} 
\begin{caption} 
{Collision of pulses. 
(a) Sine functions of the angles $\alpha =\vartheta $  
and $\varphi $ versus  $w/w_0$ at $\zeta =0$ (the boundary conditions).  
(b) The same as in (a) for $\zeta =10w_0$ (numerical simulation results). 
(c) The same as in (b) for 
$\zeta =30w_0$. See the text for more detailed description.}
\end{caption} 
\label{fig3} 
\end{figure} 

\section{Particular regimes of propagation}

There are a few particular regimes of adiabatic pulse propagation 
admitting analytic treatment. The first one occurs if atoms are 
prepared initially in a statistical mixture described by the density 
matrix $\hat{\varrho }=\frac 12\left( \left| \Phi ^{(1)} \right \rangle 
\left \langle \Phi ^{(1)} \right| + \left| \Phi ^{(2)} \right \rangle 
\left \langle \Phi ^{(2)} \right| \right) $, rather than in a pure state. 
Such a mixed state remains invariant under the action of action of the 
slowly varying electromagnetic fields: $\hat{B}\hat{\varrho }\hat{B}^\dag 
=\hat{\varrho }$, where $\hat{B}$ is given by Eq.~(\ref{B2nu}). 
Statistical averaging over 
$\hat{\varrho }$ is equivalent to averaging over the parameter $\beta $ 
uniformly distributed between 0 and $2\pi $, without any correlation 
with the instantaneous values of $\varphi $ and $\vartheta $. The result 
of statistical averaging of Eqs.~(\ref{trzve}) is 
\begin{equation} 
\frac {\partial \varphi }{\partial \zeta }=-\frac 12 
\frac {\partial \varphi }{\partial w }     , \qquad 
\frac {\partial \vartheta }{\partial \zeta }=-\frac 12 
\frac {\partial \vartheta }{\partial w }   .  \label{sx}
\end{equation} 
Equations for $\varphi $ and $\vartheta $ become decoupled. Their 
solution $\varphi =\varphi _0(w-\frac 12\zeta )$, $\vartheta =
\vartheta _0(w-\frac 12\zeta )$ describes independent propagation 
of perturbations of $\vartheta $ and $\varphi $ at the same 
group velocity $v_g=\left[ c^{-1}+G/(2\Omega ^2)\right] ^{-1}$.

\begin{figure} 
\begin{center} 
\centerline{\epsfig{file=bt4.ps,width=7cm}}
\end{center} 
\vspace{-0.6cm} 
\begin{caption} 
{Adiabatic pulse propagation for the particular initial 
condition $\beta =0$. 
(a) Sine functions of the angles $\alpha =\vartheta $  
and $\varphi $ versus  $w/w_0$ at $\zeta =0$ (the boundary conditions).  
(b) The same as in (a) for $\zeta =20w_0$ (numerical simulation results).}
\end{caption} 
\label{fig4} 
\end{figure} 

We may hazard  a conjecture 
what occurs if atoms are prepared in a mixed state with 
the density matrix $\hat{\varrho }^\prime 
=q \left| \Phi ^{(1)} \right \rangle 
\left \langle \Phi ^{(1)} \right| +(1-q)\left| \Phi ^{(2)} \right \rangle 
\left \langle \Phi ^{(2)} \right|  $, $0\le q\le 1$. It is likely 
that there are always two classes of pulses with well defined group 
velocities. If $q$ grows from 0 to 0.5, one of these velocities 
decreases whereas the other increases. At $q=0.5$ they achieve the 
same value mentioned in the previous paragraph, and then again 
restore their values $c$ and $(c^{-1}+G/\Omega ^2)^{-1}$, as $q$ 
approaches 1. At least, it can be proven easily in the perturbative 
regime, when the changes of both $\varphi $ and $\vartheta $ are small. 

Another interesting regime is related to particular initial 
conditions $\beta =0$ or $\beta =\pi /2$. Let all the atoms be pumped 
initially into the state $\left| 3\right \rangle $. The fields are 
switched on in the following order, which is  
a generalization of the counterintuitive pulse order for a 
$\Lambda $-medium \cite{adiabaton}: Initially, at $w=0$ only the field 
driving the empty transition $\left| 2\right \rangle \leftrightarrow 
\left| 0\right \rangle $ is present, i.e., $\varphi =\vartheta =0$. 
Obviously, $\beta =0$. 
Then the field driving the transition $\left| 1\right \rangle 
\leftrightarrow \left| 0\right \rangle $ is switched on adiabatically, so  
that $\vartheta $ grows and then is kept constant at a certain level. 
Finally,  the field driving the transition $\left| 3\right \rangle 
\leftrightarrow \left| 0\right \rangle $ is switched on.  

When $\vartheta $ changes, $\sin \varphi =0$. Then, according to 
Eq.~(\ref{basenu}), $\nu $ remains zero, and Eqs.~(\ref{workn}) are 
reduced to $\partial \varphi /(\partial u_1)=0$, $\partial \vartheta /
(\partial u_2)=0$. Such a propagation regime occurs unless the front 
of the $\vartheta $-pulse, propagating at the slow group velocity, 
approaches the front of the $\varphi $-pulse, propagating at $c$. 

Thus one has a possibility of preparation of a tripod medium in any 
desired coherent superposition of low-energy states. Numerical results 
presented in Fig.~3 illustrate this conclusion: Finally, atoms in the 
region $0<z<20 w_0$ are prepared in the state $-0.29\left| 1\right \rangle 
-0.53 \left| 2\right \rangle -0.80\left| 3\right \rangle $, as can be 
derived from the values $\varphi =-0.65$, $\vartheta =0.50$ at $w=40w_0$. 
Then one can suddenly change the laser radiation parameters in such a 
manner that this state will correspond to a coherent superposition of the 
two dark states defined with respect to the new values of the Rabi 
frequencies, thus obtaining a new value for the parameter $\beta $. 

The case of $\beta =\pi /2$ is physically equivalent to the 
previous one, differing only in notation of the states and 
electromagnetic fields.  

\section{Conclusion} 

Requirements for experimental implementation of adiabatic pulse 
propagation in a tripod medium should not differ from that for slow 
light propagation in $\Lambda $-media \cite{harrev,glc,lrmp}. 
A method for initial preparation of a tripod medium in any desired 
superposition state was outlined in the previous section. 
For example, consider a tripod medium with the following 
parameters: $d_{0j}\approx 10^{-18}$~esu$\, \cdot \, $cm, $k_j\approx 
10^5$~cm$^{-1}$, $n\approx 10^{12}$~cm$^{-3}$. Let the total laser 
intensity be of about 3~mW/cm$^2$ (slightly below the atomic transition 
saturation limit). Hence, $\Omega \approx 2.4\cdot 10^6$~s$^{-1}$ and 
$\Omega ^2/G\approx 10^4$~cm/s (i.e., the group velocity of the slow 
pulses is $v_g\approx 0.3\cdot 10^{-6}c$). The 
value of the scaling parameter of the 
horizontal axes of the Figs.~2 -- 4 $w_0\approx 0.1$~cm is thus large  
enough to provide adiabaticity. Therefore the processes illustrated in 
Fig.~2--4 can be observed in a few centimeter long gas cell. The time 
delay between the fast and slow pulses is of about 0.3~ms, therefore 
the lifetime of 
coherence between the states $\left| j\right \rangle $, $j=1,2,3$, 
should be 1~ms or longer. It is achievable in coated cells or cells 
with a buffer gas.  

To conclude, we have investigated electromagnetic pulse propagation in a 
coherent atomic medium with the tripod configuration of levels in the 
adiabatic regime. The propagation equations (\ref{wa1} -- \ref{wa3}) are 
derived and their solutions in the form of slow [Eq.~(\ref{slow1})] and 
fast [Eq.~(\ref{fast1})] pulses are obtained analytically. Our 
numerical simulations confirm that these solutions are general 
asymptotic solutions for any incident pulse of a finite duration. 
We have suggested a method of preparation of a tripod medium in an 
arbitrary superposition of the low-energy states based on switching 
on the laser fields in a counterintuitive order. The tripod scheme 
provides two novel features in comparison to the $\Lambda $-scheme. 
The first one is adiabatic pulse propagation in a medium prepared in a 
statistical mixture of the two dark states. The second one is the 
possibility of collisions between the slow and fast pulses revealing that 
they change their shapes after nonlinear interaction and thus do not 
satisfy the classical definition of a soliton \cite{soliton1}. 

The author thanks L.~Windholz, E.V.~Galaktionov, 
and D.G.~Yakovlev for useful discussions. 
The work is supported by the Austrian Science Foundation (project 
P~14645) and the program Leading Russian Scientific Schools (grant 
1115.2003.2).


\begin{thebibliography}{99}
\bibitem{arimrev} E. Arimondo, in:\ E.\ Wolf (Ed.), Progress in Optics, vol.
35, North-Holland, Amsterdam, 1996, p.259.

\bibitem{harrev} S.E. Harris, Phys. Today {\bf 50}, No. 7, 36 (1997). 

\bibitem{adiabaton} R. Grobe, F.T. Hioe, J.H. Eberly, Phys. Rev. Lett. 
{\bf 73
}, 3183 (1994).

\bibitem{glc} A. Kasapi, M. Jain, G. Y. Yin, S. E. Harris, Phys. Rev. Lett.
{\bf 74}, 2447  (1995); 
O. Schmidt, R. Wynands, Z. Hussein, D. Meschede, Phys. Rev.
A {\bf 53}, R27 (1996); 
L.V. Hau, S.E. Harris, Z. Dutton, C.H. Behroozi, Nature {\bf 397}, 594 
(1999); M.M. Kash, V.A. Sautenkov, A.S. Zibrov, L. Hollberg, G.R. Welch,
M.D. Lukin, Yu. Rostovtsev, E.S. Fry, M.O. Scully, Phys. Rev. Lett. 
{\bf 82}, 5229 (1999); D. Budker, D.F. Kimball, S.M. Rochester, 
V.V. Yashchuk, Phys.
 Rev. Lett. 
{\bf 83}, 1767 (1999); E. Podivilov, B. Sturman, A. Shumelyuk, 
S.
 Odoulov, Phys. Rev. Lett. {\bf 91}, 083902 (2003).

\bibitem{harhau} S.E. Harris, L.V. Hau, Phys. Rev. Lett. {\bf 82}, 
4611 (1999).

\bibitem{kerr} H. Kang, Y. Zhu, Phys. Rev. Lett. {\bf 91}, 
093601 (2003).

\bibitem{switch} D.A. Braje, V. Bali\'{c}, G.Y. Yin, S.E. Harris, Phys. Rev.
A {\bf 68}, 041801 (2003).

\bibitem{lrmp} M.D. Lukin, Pev. Mod. Phys. {\bf 75}, 457 (2003).

\bibitem{bgap} A.\ Andr\'{e}, M.D. Lukin, Phys. Rev. Lett. {\bf 89}, 
143602 (2002);
 M. Bajcsy, A.S. Zibrov, M.D. Lukin, Nature, {\bf 426}, 
638 (2003).

\bibitem{bergth} R. Unanyan, M. Fleischhauer, B.W. Shore, K. Bergmann, Opt.
Commun. {\bf 155}, 144 (1998); R. Unanyan, B.W. Shore, K. Bergmann, 
Phys. Rev. A {\bf 59}, 2910 
(1999).

\bibitem{bergex} H. Theuer, R.G. Unanyan, C. Habscheid, K. Klein, K.
Bergmann, Opt. Express {\bf 4}, 77 (1999).

\bibitem{renzoni} Z. Kis, F. Renzoni, Phys. Rev. A {\bf 65},  032318 
(2002).

\bibitem{stenholm} Z. Kis, S. Stenholm, Phys. Rev. A {\bf 64},  063406 
(2001).

\bibitem{wilcz} F. Wilczek, A. Zee, Phys. Rev. Lett. {\bf 52}, 2111 (1984).

\bibitem{berry} M.V. Berry, Proc. R. Soc. London, Ser. A {\bf 392}, 45 
(1984).

\bibitem{pk} E. Paspalakis, P.L. Knight, J. Opt. B: Quant. Semiclass. Opt. 
{\bf 4},  S372 (2002).

\bibitem{pm} D. Petrosyan, Yu.P. Malakyan, e-print: quant-ph/0402070.

\bibitem{hamb} B.S. Ham, P.R. Hemmer, Phys. Rev. Lett. {\bf 84}, 
4080 (2000).

\bibitem{hama} B.S. Ham, Appl. Phys. Lett. {\bf 78}, 3382 (2001).

\bibitem{neos} I.E. Mazets and B.G. Matisov, Quant. Semiclass. Opt.  
{\bf 8}, 909 (1996). 

\bibitem{ds} I.E. Mazets, Phys. Rev. A {\bf 54}, 3539 (1996). 

\bibitem{soliton1} N.J. Zabusky and M.D. Kruskal, Phys. Rev. {\bf 15}, 
240 (1965). 
\end{thebibliography}
\end{document}